%% file: ISMIR2026_template_cameraready.tex
\documentclass{article}
\usepackage[T1]{fontenc}
\usepackage{microtype}
\usepackage[utf8]{inputenc}
\usepackage[]{ismir} %
\usepackage{amsmath,amssymb,cite,url}
\usepackage{graphicx}
\usepackage{color}
\usepackage{booktabs}
\usepackage{tikz}
\usetikzlibrary{arrows.meta,positioning}
\usepackage{siunitx}

\usepackage[nolist]{acronym}

\begin{acronym}
\acro{ldm}[LDM]{Latent Diffusion Model}
\acro{mmse}[MMSE]{Minimum Mean Squared Error}
\acro{dit}[DiT]{Diffusion Transformer}
\acro{kad}[KAD]{Kernel Audio Distance}
\acro{ae}[AE]{Autoencoder}
\acro{cd}[CD]{cosine distance}
\acro{srs}[SRS]{Smule Renaissance Small}
\acro{m2l}[M2L]{Music2Latent}
\acro{em}[EM]{Expectation Maximization}
\end{acronym}

\title{ Music Restoration via Latent Operator Optimization and Diffusion Model Priors}

\multauthor
  {Michal Švento$^1$ \hspace{1cm} Eloi Moliner$^2$ \hspace{1cm} Valtteri Kallinen$^2$}
  {{\bf Lauri Juvela$^2$ \hspace{1cm} Vesa Välimäki$^2$ \hspace{1cm} Pavel Rajmic$^1$}\\
  $^1$ Dept.\ of Telecommunications, Brno University of Technology, Czech Republic\\
  $^2$ Acoustic Lab, Dept. of Information and Communications Engineering, Aalto University, Finland\\
  {\tt\small michal.svento@vut.cz, eloi.moliner@aalto.fi}
  }

\def\authorname{M. Švento, E. Moliner, V. Kallinen, L. Juvela, V. Välimäki and P. Rajmic}

\usepackage[bookmarks=false,pdfauthor={\authorname},pdfsubject={\pdfsubject},hidelinks]{hyperref}

\begin{document}

\maketitle

\begin{abstract}
Music restoration seeks to recover a clean signal from an observed recording degraded by an unknown effect, distortion, or corruption.
Existing systems often rely on paired training data and distortion-specific supervision,
which limits their use when the forward process is not known in advance.
We propose LOUDAR (Latent-space Optimization of Unknown Distortion for Audio Restoration) a general-purpose restoration method that operates in the latent space of a pretrained audio autoencoder and models the unknown distortion as a learnable latent operator. 
At inference time, LOUDAR alternates between estimating the clean latent variable and
updating the latent operator parameters.
An unconditional latent diffusion model provides a prior over clean audio and regularizes this inference by steering the latent estimate toward the manifold of clean recordings. 
Because the degradation model is adapted per input, 
the approach is broadly applicable across diverse restoration problems.
We evaluate LOUDAR on singing voice effect removal and restoration,
as well as guitar distortion removal, 
and show that it consistently improves over degraded inputs and is competitive with supervised and unsupervised baselines in waveform and latent domains.
\end{abstract}

\section{Introduction}\label{sec:introduction}
\vspace{0pt}

Music recordings are rarely observed in a dry or studio-ready form.
Vocals, guitars, and full mixes are typically shaped by chains of effects such as reverberation, compression, and saturation.
Recordings may also be affected by distortions such as compression artifacts, band-limitation or clipping.
In many cases, 
it is desirable to recover the 
unprocessed signal from such observations, 
a process we refer to as music ``restoration.'' 
Such recovery should ideally remain possible even when the degradations or effect chains, which we broadly refer to as ``distortion,'' are unknown and overly complex. 
This capability is useful for applications such as music production, remixing or editing, as well as for preparing training data for music technology systems such as automatic mixing \cite{martinez2022automatic}, large language model-guided post-production \cite{doh2025llm2fx}, or source separation \cite{zang2025music}.

Most restoration approaches rely on supervised training with paired clean/distorted data and a predefined family of distortions assumed during model development \cite{imort2022distortion,rice2023general,take2024audio,li2025apollo}. 
This assumption is limiting in practice. 
Audio effects applied to music can be highly complex
in serial and parallel processing chains \cite{lee2023blind}, 
so designing a comprehensive 
pipeline 
is nearly impossible. 
Moreover, certain corruption processes are inherently difficult to simulate and lack a clean reference entirely, 
such as historical recordings \cite{moliner2024diffusion}.
As a result, supervised systems are tailored to the distortions represented in training and can struggle with more complex or mismatched effect chains at test time \cite{lemercier2025unsupervised}. 

There is, nonetheless, an unsupervised alternative: framing restoration as a blind inverse problem, where a diffusion model trained on clean audio serves as a prior while jointly estimating a parametric distortion model for approximate posterior sampling.
Existing methods in this paradigm, however, typically remain restricted to particular forward-model families \cite{lemercier2025unsupervised,moliner2024diffusion,moliner2025unsupervised,xu2025arraydps}.
 A general restoration framework that can adapt at test time to an unknown 
 processing chain while exploiting strong priors over clean music is still lacking.

Learned audio representations provide an appealing foundation for such a framework ~\cite{liu23-audio-ldm, evans2024-stable-audio}. \acp{ae} map waveforms to compact and semantically-organized latent spaces.
Prior work suggests latent-space restoration as a technically promising direction \cite{wang2023audit,bralios2025learning,dhyani2025high,raphaeli2025silo}. 
We hypothesize that such representations enable tractable unsupervised restoration in latent space 
by compressing both the representation and the distortion model,
enabling a compact neural network to approximate a broad range of distortions while remaining easy to optimize.

This paper introduces LOUDAR (Latent-space Optimization of Unknown Distortion for Audio Restoration), an unsupervised framework for music restoration.
LOUDAR follows a similar inference strategy as recent unsupervised restoration methods \cite{lemercier2025unsupervised,moliner2025unsupervised}, but transfers it to the latent space of an \ac{ae}.
The proposed method parameterizes the unknown distortion directly in latent space using a learnable operator,
while an unconditional \ac{ldm} trained on clean music provides a prior over unprocessed signals.
At test time, LOUDAR alternates between estimating a clean latent vector and updating the latent distortion operator and residual perturbation, yielding an \ac{em} inference procedure specialized to the input recording.

Our main contributions are threefold: (i) a formulation of music restoration as a blind inverse problem in the latent space; (ii) a compact, general-purpose latent operator architecture designed to stabilize blind optimization; and (iii) a comprehensive evaluation across singing voice and guitar domains, using objective and subjective metrics.
Code and listening examples are available on the project website\footnote{\url{https://michalsvento.github.io/loudar/}}.

\vspace{-5pt}
\section{Methods}  \label{sec:methods}
\vspace{-3pt}

\subsection{Problem Definition}
\vspace{-2pt}
\label{sec:problem-def}

We consider restoration problems where an observed waveform is generated from a clean one through an unknown distortion process. 
Let $\mathbf{x} \in \mathbb{R}^L$ and $\mathbf{y} \in \mathbb{R}^L$ denote the clean and distorted waveforms, respectively, both of length $L$.
We assume the forward process
\begin{equation}
\mathbf{y} = f(\mathbf{x}) + \mathbf{n},
\end{equation}
where $f \colon \mathbb{R}^L \to \mathbb{R}^L$ is an unknown, possibly nonlinear,
distortion operator and $\mathbf{n} \in \mathbb{R}^L$ is an additive residual term.
This process involving waveforms is shown in the top part of Figure~\ref{fig:latent-operator}. 
Our goal is to recover the clean (unprocessed) waveform $\mathbf{x}$ given an observation $\mathbf{y}$.

Inspired by \cite{raphaeli2025silo}, rather than solving this inverse problem in the waveform domain,
we work in the latent space of a pretrained (frozen) audio \ac{ae}, with encoder $E \colon \mathbb{R}^L \to \mathbb{R}^{C \times N}$ and decoder $D \colon \mathbb{R}^{C \times N} \to \mathbb{R}^L$\!.
The clean and distorted latent vectors are denoted by $\mathbf{z}_0 := E(\mathbf{x}) \in \mathbb{R}^{C \times N}$ and $\mathbf{z}_y := E(\mathbf{y}) \in \mathbb{R}^{C \times N}$, respectively, where 
$C$ is the channel size and 
$N$ is the number of latent frames.
We denote $p_x$ the clean waveform data distribution, and $p_z$ the corresponding clean latent distribution induced by the encoder $E$.
Empirical access to these distributions is provided by a collection of clean training examples $\{\mathbf{x}^i\}_i \sim p_x$ and corresponding latent vectors $\{\mathbf{z}_0^i\}_i = \{E(\mathbf{x}^i)\}_i \sim p_z$.
The \ac{ae} is also assumed to approximately reconstruct clean signals on the support of $p_x$, i.e., $D(E(\mathbf{x})) \approx \mathbf{x}$ for $\mathbf{x} \sim p_x$.

The distorted latent vector is then modeled as
\begin{equation}
\mathbf{z}_y = E\bigl(f(\mathbf{x}) + \mathbf{n}\bigr) \approx g_\phi(\mathbf{z}_0) + \mathbf{r},
\end{equation}
where $g_{\boldsymbol{\phi}} \colon \mathbb{R}^{C \times N} \to \mathbb{R}^{C \times N}$ is a parametric latent distortion operator with unknown parameters ${\boldsymbol{\phi}}$.
The residual 
$\mathbf{r} \in \mathbb{R}^{C \times N}$
relaxes the time-invariance of $g_\phi$, absorbing small localized artifacts or noise it cannot represent.
Figure~\ref{fig:latent-operator} illustrates this signal model.
While \cite{raphaeli2025silo} assume knowledge of $g_{\boldsymbol{\phi}}$ by pre-training the latent operator in a dedicated stage using paired data, we treat this setup as a blind inverse problem in the latent space:
given only $\mathbf{z}_y$, estimate the clean latent code $\mathbf{z}_0 \sim p_z$ together with the operator 
parameters ${\boldsymbol{\phi}}$ and the residual matrix $\mathbf{r}$.

\input{figures/latent_operator_figure}

\vspace{0pt}
\subsection{Latent Diffusion Model Prior}
\vspace{0pt}
\label{sec:ldm}

\label{sec:latent_diffusion}

We use an unconditional \ac{ldm} as a prior over the clean latent distribution $p(\mathbf{z})$.
Diffusion models define a simple forward process that gradually corrupts samples from the target distribution $p(\mathbf{z})$ with Gaussian noise, and then learn to reverse this process by denoising \cite{ho2020denoising}.
 Given a clean latent vector $\mathbf{z}_0 \sim p_z$ and noise $\boldsymbol{\varepsilon} \sim \mathcal{N}(\mathbf{0}, \mathbf{I})$, the forward perturbation at any noise level $\tau \in [0, \sigma_{\mathrm{max}}]$ is defined as
\begin{equation}
    \mathbf{z}_{\tau} = \mathbf{z}_0 + \tau\boldsymbol{\varepsilon}.
\end{equation}
Sampling from $p(\mathbf{z})$ is then performed by approximately reversing this process: starting from $\mathbf{z}_{\sigma_{\mathrm{max}}} \sim \mathcal{N}(\mathbf{0}, \sigma^2_{\mathrm{max}} I)$, the model progressively denoises toward $\tau = 0$.
In the framework proposed by Karras et al. \cite{Karras2022edm}, the continuous reverse-time dynamics are characterized by the probability flow ordinary differential equation (PF-ODE):
\begin{equation}
\label{eq:pf-ode}
    \textup{d}\mathbf{z}_{\tau} = -\tau \nabla_{\mathbf{z}_{\tau}}\log p(\mathbf{z}_{\tau}) \, \textup{d}\tau.
\end{equation}
The key quantity is the score function $\nabla_{\mathbf{z}_{\tau}}\log p(\mathbf{z}_{\tau})$.
At each noise level, it points toward regions of higher $p_{\mathbf{z}_{\tau}}$ probability density. 
The score function is generally intractable, but it can be approximated with a neural network $s_\theta(\mathbf{z}_\tau, \tau) \approx \nabla_{\mathbf{z}_{\tau}}\log p(\mathbf{z}_{\tau})$ with parameters $\theta$.

A useful property of Gaussian perturbations is that the score function is directly related to the Minimum Mean Square Error denoiser $\mathbb{E}[\mathbf{z}_0 \mid \mathbf{z}_\tau]$.
We denote its learned approximation by $\hat{\mathbf{z}}_{\theta}(\mathbf{z}_{\tau}, \tau) \approx \mathbb{E}[\mathbf{z}_0 \mid \mathbf{z}_\tau]$, a deep neural network with parameters $\theta$.
The score follows directly as
\begin{equation}
    \nabla_{\mathbf{z}_{\tau}} \log p(\mathbf{z}_{\tau})
    \approx s_{\theta}(\mathbf{z}_{\tau}, \tau)
    = \frac{\hat{\mathbf{z}}_{\theta}(\mathbf{z}_{\tau}, \tau) - \mathbf{z}_{\tau}}{\tau^2}.
\end{equation}
In our setup, the denoiser is parameterized as 
\begin{equation}\label{eq:denoiser}
    \hat{\mathbf{z}}_{\theta}(\mathbf{z}_{\tau}, \tau) = c_{\text{skip}}(\tau)\mathbf{z}_{\tau} + c_{\text{out}}(\tau)F_{\theta}(c_{\text{in}}(\tau)\mathbf{z}_{\tau}, \tau),
\end{equation}
where the scaling coefficients are defined as $c_{\text{in}}(\tau) = c_{\text{skip}}(\tau) = 1/(\tau+1)$ and $c_{\text{out}}(\tau) = \tau/(\tau+1)$, 
and $F_\theta$ a \ac{dit} backbone \cite{peebles2023scalable}.
We train $F_\theta$ with the Rectified Flow objective \cite{liu2022flow}:
\begin{equation}
    \label{eq:rf-objective}
    \mathcal{L}(\theta) = \mathbb{E}_{\mathbf{z}_0, \epsilon, \tau} \left[ \| F_{\theta}(c_{\text{in}}(\tau)\mathbf{z}_{\tau}, \tau) - (\boldsymbol{\varepsilon} - \mathbf{z}_0) \|_F^2 \right].
\end{equation}

\vspace{0pt}
\subsection{Diffusion Posterior Sampling}
\vspace{0pt}
\label{sec:inference_algorithm}

Diffusion models can also be used to solve inverse problems at test time, without additional training, and several strategies have been proposed \cite{daras2024survey}. In this work, we adopt Diffusion Posterior Sampling (DPS) \cite{chung2023dps}, which targets the posterior distribution $p(\mathbf{z}_0 \mid \mathbf{z}_y)$.

 By Bayes' rule, the posterior score decomposes into a~prior term and a likelihood term:
\begin{equation}
\label{eq:conditional-score}
    \nabla_{\mathbf{z}_{\tau}} \log p(\mathbf{z}_{\tau} \mid \mathbf{z}_y)
    = \nabla_{\mathbf{z}_{\tau}}\! \log p(\mathbf{z}_{\tau})
    + \nabla_{\mathbf{z}_{\tau}}\! \log p(\mathbf{z}_y \mid \mathbf{z}_{\tau}).
\end{equation}
The first term is the unconditional score, modeled by the diffusion prior $s_{\boldsymbol{\theta}}(\mathbf{z}_{\tau}, \tau)$, which steers samples toward the clean latent manifold. The second term is a likelihood score that guides the reverse process toward posterior samples that remain consistent with the observation $\mathbf{z}_y$~\cite{chung2023dps}.
Although the likelihood score is generally intractable, it can be approximated using the forward model as $\nabla_{\mathbf{z}_{\tau}} \log p(\mathbf{z}_y \mid \mathbf{z}_{\tau}) \approx l(\mathbf{z}_y, \mathbf{z}_{\tau}; {\boldsymbol{\phi}}, \mathbf{r})$. We use the following approximation, similar to \cite{chung2023dps}:
\begin{equation}\label{eq:likelihood}
    \hspace{-2pt}
    l(\mathbf{z}_y, \mathbf{z}_{\tau}; {\boldsymbol{\phi}}, \mathbf{r}) 
    \hspace{-2pt}
    \approx 
    \hspace{-2pt}
    -
    \zeta_{\tau} \nabla_{\mathbf{z}_{\tau}} \|\mathbf{z}_y 
    \hspace{-1pt}
    -
    \hspace{-1pt}
    \left(g_{{\boldsymbol{\phi}}}(\hat{\mathbf{z}}_{\boldsymbol{\theta}}(\mathbf{z}_{\tau}, \tau)) 
    \hspace{-2pt}
    + 
    \hspace{-2pt}
    \mathbf{r}\right)\|_F^2,
\end{equation}
where $\zeta_\tau$ is a noise-level dependent scaling factor.
However, in our setting the latent operator parameters $\boldsymbol{\phi}$ and the residual term $\mathbf{r}$ are unknown. We therefore infer them jointly with the clean latent through an alternating blind inference procedure, described next.

\vspace{0pt}
\subsection{Distortion-Blind Inference}
\vspace{0pt}

We adopt an EM-style alternating inference algorithm for blind inverse problems along the reverse diffusion trajectory, following prior work on inverse problems \cite{laroche2024fast, lemercier2025unsupervised, moliner2025unsupervised}.
At a high level, the goal is to estimate the unknown operator parameters by maximizing the expected log-likelihood of the observation under the posterior of the clean latent:
\begin{equation}
    (\boldsymbol{\phi}, \mathbf{r}) = \arg \max_{\boldsymbol{\phi}, \mathbf{r}}\
    \mathbb{E}_{p(\mathbf{z}_0 \mid \mathbf{z}_y)} \!\left[\log p(\mathbf{z}_y \mid \mathbf{z}_0;\, \boldsymbol{\phi}, \mathbf{r})\right].
\end{equation}
In practice, the inference alternates between three steps, which are also shown in Figure~\ref{fig:inference-algorithm}.
At each reverse-time step, we estimate the clean latent under the current operator (E-step), update the operator to better explain the observation (M-step), and then take one guided ODE step (update).

\input{figures/inference_figure}

\textbf{E-step:} At iteration $i$ (running from $M$ down to $0$), we compute the clean-latent estimate via \eqref{eq:denoiser}: $\hat{\mathbf{z}}_0^{(i)} := \hat{\mathbf{z}}_{\boldsymbol{\theta}}(\mathbf{z}_{\tau_i}, \tau_i)$.
Treating $\hat{\mathbf{z}}_0^{(i)}$ as a Dirac approximation to the intractable posterior $p(\mathbf{z}_0 \mid \mathbf{z}_y)$ yields the surrogate log-likelihood:
\begin{equation}
    \mathbb{E}_{p(\mathbf{z}_x \mid \mathbf{z}_y)} \!\left[\log p(\mathbf{z}_y 
    \hspace{-2pt}
    \mid
    \hspace{-2pt}
    \mathbf{z}_0; \boldsymbol{\phi}_i, \mathbf{r}_i)\right] 
    \approx \log p\!\left(\mathbf{z}_y 
    \hspace{-2pt}
    \mid
    \hspace{-2pt}
    \hat{\mathbf{z}}_0^{(i)}; \boldsymbol{\phi}_i, \mathbf{r}_i\right),
\end{equation}
reducing the E-step to a single forward-pass evaluation rather than an explicit posterior sampling.

\textbf{M-step:} With fixed estimate $\hat{\mathbf{z}}_0^{(i)}$\!,
we update the operator parameters and residual perturbation by optimizing:
\begin{equation} \label{eq:mstep}
    \begin{aligned}
        (\boldsymbol{\phi}_{i+1}, \mathbf{r}_{i+1}) = \arg \min_{\phi, \mathbf{r}}\ 
        &\|\mathbf{z}_y - (g_{\boldsymbol{\phi}}(\hat{\mathbf{z}}_0^{(i)}) + \mathbf{r})\|_1 \\
        &+ \gamma_{\boldsymbol{\phi}} \| \boldsymbol{\phi} \|_2^2 + \gamma_\mathbf{r} \| \mathbf{r} \|_F^2.
    \end{aligned}
\end{equation}
In practice, we optimize this objective with AdamW \cite{loshchilov2018decoupled} and apply decoupled weight decay to both $\boldsymbol{\phi}$ and $\mathbf{r}$; the quadratic penalties above summarize this regularization.

\textbf{Update:} The update towards the clean latent variable is an Euler step along the reverse dynamics:
\begin{equation}
    \mathbf{z}_{\tau_{i-1}} =
    \mathbf{z}_{\tau_i} - 
    \Delta_{\tau_i}
    [s_{\theta}(z_{\tau_i}, \tau_i) + l(\mathbf{z}_y, \mathbf{z}_{\tau_i}; \boldsymbol{\phi}_{i+1}, \mathbf{r}_{i+1})],
\end{equation}
where $\Delta_{\tau_i}=(\tau_i - \tau_{i-1})$ denotes the step size.

Alternating these steps adapts the latent operator and residual
while steering the trajectory toward the clean latent manifold.
The restored signal is obtained via the decoder, $\hat{\mathbf{x}} = D(\hat{\mathbf{z}}_{\theta}{(\mathbf{z}_{\tau_0}}, \tau_0))$, back to the waveform domain, as shown at the bottom of Figure \ref{fig:inference-algorithm}.

\vspace{0pt}
\subsection{Latent Operator Model}
\vspace{0pt}
\label{sec:latent_op}

\input{figures/latent_operator_arch_figure}

A crucial design choice in LOUDAR is how to parameterize the unknown latent operator $g_{\boldsymbol{\phi}}$.
The model should be flexible enough to represent a broad range of distortions, while remaining sufficiently constrained to avoid overfitting during each M-step optimization.
This trade-off is especially important because blind optimization is performed along the reverse diffusion trajectory, where the intermediate clean-latent estimate $\hat{\mathbf{z}}_0$ can be inaccurate at early steps.
Without explicit capacity control, the operator can overfit these transient estimates and destabilize the subsequent denoising updates.

We implement $g_{\boldsymbol{\phi}}$ as a causal convolutional network acting on latent matrices.
Figure~\ref{fig:latent-operator-arch} summarizes the architecture.
Convolutions along the temporal axis preserve translation equivariance over time, which matches many audio effects and distortions, while the temporal compression of the latent representation makes even a small receptive field correspond to a longer waveform context.
In the default configuration used throughout this work, $g_{\boldsymbol{\phi}}$ contains three bias-enabled low-rank causal Conv1D layers with kernel size $3$ and channel size $64$.
The first two layers are followed by ReLU nonlinearities, and residual connections are applied at every layer.
To limit capacity, each layer is implemented through a rank-$12$ intermediate channel space: instead of applying the temporal convolution directly at the full 64-channel dimensionality, the layer projects to 12 channels, performs the causal convolution in that reduced space, and projects back to 64 channels. The projection layers are denoted as $U$ and $V$ in Figure \ref{fig:latent-operator-arch}.

The additive residual term $\mathbf{r}$ complements $g_{\boldsymbol{\phi}}$ by capturing sample-specific mismatch that is not well explained by the shared convolutional operator.
Unlike $g_{\boldsymbol{\phi}}$, however, $\mathbf{r}$ has no temporal structure, so it is more prone to absorbing noise or other interferences.
Consequently, we treat $\mathbf{r}$ as a~free optimizable matrix and apply a weight penalty to it, as well as to the weights $\boldsymbol{\phi}$, as described in \eqref{eq:mstep}.

\vspace{0pt}
\section{Experiments and Results}\label{sec:experiments-draft}
\vspace{0pt}

We evaluate LOUDAR on singing voice and guitar to test whether the same method transfers across substantially different musical material and remains competitive with representative baselines. For singing voice, we study removal of common vocal production effects (objective and subjective evaluation) and restoration from nonlinear distortions (subjective only). For guitar, we study recovery of the  direct injection (DI) signal from amplifier-processed recordings.

\vspace{0pt}
\subsection{Experimental Details}
\vspace{0pt}

\noindent\textbf{Autoencoder.}
We adopt \ac{m2l} \cite{pasini2024m2l}, an \ac{ae} based on a consistency model \cite{song2023consistency},
as the latent representation employed in our experiments.
This \ac{ae} encodes 48~kHz audio waveforms to a temporal rate of 12~Hz with a channel size of 64, corresponding to a dimensionality reduction of approximately $64\times$.
In preliminary experiments, \ac{m2l} proved qualitatively more effective for our specific task than alternative variational \acp{ae} \cite{evans2024-stable-audio}.
We use the \ac{m2l} weights released by%
~\cite{torres2026learning},
obtained with a larger training dataset.
In preliminary experiments, \ac{m2l} proved qualitatively more effective than alternative Variational AEs \cite{evans2024-stable-audio}, which we attribute to its lower dimensionality, favorable diffusability properties \cite{skorokhodov2025improving}, and decoder robustness.

\noindent\textbf{Diffusion Model.}
The diffusion backbone, $F_{\theta}$, is implemented as an unconditional Diffusion Transformer (DiT) \cite{peebles2023scalable}. The implementation is based on Stable Audio Open \cite{evans2025sao} and contains approximately $68$\,M trainable parameters.
All models operate on monaural audio segments of $526,080$ samples (approximately $11$\,s at $48$ kHz). In the M2L representation, these segments correspond to $128 \times 64$ matrices.
Models are trained by minimizing the objective in \eqref{eq:rf-objective} using the AdamW optimizer \cite{loshchilov2018decoupled} with a learning rate of $10^{-4}$\!, weight decay of $0.01$, $(\beta_1, \beta_2) = (0.9, 0.999)$, and a~batch size of 16. The training data varies by experimental condition, as detailed in the sections below.

\noindent\textbf{Inference Parameters.}
To ensure robustness across all experiments, we maintain a consistent hyperparameter configuration for our inference algorithm. We utilize $M = 300$ steps following the reverse-time schedule proposed in EDM
\cite{Karras2022edm}, with boundary noise levels $\sigma_{\mathrm{max}} = 30$, $\sigma_{\mathrm{min}} = 10^{-4}$\!, and a curvature parameter $\rho = 7$. Furthermore, we incorporate stochasticity into the updates as described in \cite{Karras2022edm} to mitigate generation artifacts in the clean estimates, setting $S_{\mathrm{churn}} = 20$.
Following \cite{moliner2023solving}, the DPS scaling factor $\zeta_\tau$ is parameterized as: $\zeta_\tau = \tilde{\zeta} \frac{\sqrt{CN}}{\tau \|G\|_F}$, where $G$ denotes the gradient of the measurement consistency term \eqref{eq:likelihood}.
We set $\tilde{\zeta} = 0.3$ 
empirically to balance fidelity against generative expressiveness: lower values lead to hallucinations, higher values over-rely on the observation and degrade quality.

The latent operator, as specified in Sec.\ \ref{sec:latent_op}, contains approximately $9.4\,\text{k}$ parameters.
For its optimization, we use AdamW \cite{loshchilov2018decoupled} with a learning rate of $10^{-4}$ and a purposedly high weight decay of $1$. Within each of the $M$ inference steps, we perform 10 operator optimization iterations.

\vspace{0pt}
\subsection{Evaluation}
\vspace{0pt}

\noindent\textbf{Objective Metrics.}
We evaluate our framework using specialized embedding spaces rather than waveform-level comparisons, as the latter are often ill-suited for assessing generative models \cite{gui2024FADforGenAI}. 
Furthermore, the \ac{ae} can introduce phase differences; while these are often perceptually negligible, they may unfairly penalize some methods in sample-accurate evaluations. 

We utilize the embeddings AFx-Rep mid \cite{steinmetz2024stito} and FxEncoder++ \cite{yeh2025fx}, both of which are explicitly trained to capture audio effect characteristics, alongside CLAP \cite{wu2023laionclap}, which has demonstrated robust sensitivity to audio effects~\cite{chu2025text2fx}. 
For singing voice tasks, we additionally incorporate a BYOL-based singer identity embedding \cite{torres2023singer} to monitor potential drift in vocal characteristics and ensure the preservation of the original performer's identity.

We report both pairwise and distributional metrics to capture different facets of performance. Pairwise performance is measured via the \ac{cd} between the embeddings of the clean reference and the model estimates, quantifying how accurately the model restores specific features for each individual input. To complement this, we evaluate the overall statistical consistency of the results using \ac{kad} \cite{chung2025kad}; while pairwise metrics focus on individual reconstruction accuracy, \ac{kad} assesses the alignment between the generated output manifold and the target ground truth distribution, ensuring the stylistic realism of the results.
Lower values indicate better performance for all reported measures, with zero optimal.

\noindent\textbf{Subjective Listening Test.}
To validate the perceptual quality of the proposed method against several baselines, we conducted a formal listening test using the webMUSHRA framework \cite{schoeffler2018webmushra}. The evaluation involved 13 volunteers recruited from the authors' institutions, 11 of whom were experienced in subjective audio evaluation. Participants performed the test remotely using headphones in quiet environments. 
Following the MUSHRA protocol, each trial included a “dry” reference and a distorted anchor alongside the test conditions,
which listeners rated on a scale from 0 to 100.
The test was divided into two parts: a four-trial session for singing voice effect removal and a 12-trial session for severe nonlinear distortions. 
The evaluated material, along with a comprehensive analysis of the results, are detailed in the following section. 
Statistical significance was assessed using the Wilcoxon signed-rank test with a significance level of $p < 0.05$.

\vspace{0pt}
\subsection{Singing Voice Effect Removal} \label{sec:singing}
\vspace{0pt}

To evaluate the performance of LOUDAR in singing voice restoration, we train a diffusion model on the OpenSinger dataset \cite{huang2021multi},
a 50-hour, high-quality multi-singer dataset of singers performing in Chinese.
We partitioned the data by singer, reserving six singers for a held-out test set, two for validation, and utilizing the remaining singers for training. The model was trained for 375\,k iterations.

\noindent\textbf{Benchmark.}
Our primary benchmark assesses the ability of the model to remove vocal effects common in modern music production. 
To simulate the effects, we utilize Diffvox \cite{yu2025diffvox}, a differentiable audio effects pipeline which includes a parametric equalizer, a dynamic range compressor, a feedback delay network reverberator, and a ping-pong delay. DiffVox is particularly well-suited for this evaluation as it encompasses both nonlinear processing and challenging long-term temporal dependencies.
We use 365 presets provided by the DiffVox authors, which represent reverse-engineered effect chains from professionally produced vocals. These presets were applied to the OpenSinger test set to create 365 unique 11-seconds-long examples, with a balanced representation of test singers. 
From this set, four random examples featuring different singers were included in the listening test.

\noindent\textbf{Baselines.}  
We compare LOUDAR against a set of representative baselines, beginning with supervised models retrained on the OpenSinger dataset. To facilitate supervised training, we applied a distortion pipeline of randomized equalization, compression, and reverb, similar to \cite{moliner2026automatic}.

The supervised baselines include: (1) a \emph{Latent Regressor}, which utilizes the same transformer architecture as LOUDAR but minimizes a Euclidean objective to predict the clean embedding $\mathbf{z}_0$ directly from the distorted $\mathbf{z}_y$; (2)~an~\emph{\ac{ldm} Conditional}, a latent diffusion transformer conditioned on $\mathbf{z}_y$ via input concatenation; and (3)~\emph{Apollo}~\cite{li2025apollo}, a waveform-domain restoration model %
designed for MP3 restoration and retrained for this task.

We further evaluate models pretrained on other datasets that are expected to solve the task. This includes \emph{RemFX}~\cite{rice2023general}, which employs a compositional restoration approach; \emph{SRS} \cite{zang2025smule}, a recent speech and singing voice restoration model trained on large curated datasets; and \emph{VoiceFixer} \cite{liu2021voicefixer}, a general restoration model specialized for speech. Finally, we compare LOUDAR against \emph{BUDDy} \cite{lemercier2025unsupervised}, an unsupervised waveform-domain model. While BUDDy uses an algorithmic framework similar to LOUDAR, its operator is primarily designed for linear room impulse responses, which may limit its effectiveness across the broader distortions present in this benchmark.

\begin{figure}[t]
    \centering
    \includegraphics[width=\columnwidth]{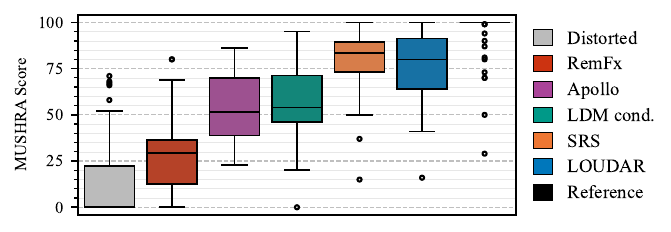}
    \vspace{-20pt}
    \caption{
    Results of the MUSHRA listening test on singing voice effect removal.
 \vspace{0pt}
    }
    \label{fig:mushra_diffvox}
\end{figure}

\noindent\textbf{Results.}
Objective metrics are reported in the upper part of Table~\ref{tab:joint_table}, while listening test results are shown in Figure~\ref{fig:mushra_diffvox}. A first observation is the apparent discrepancy between objective and perceptual evaluation. Apollo achieves the best overall scores across objective metrics (see Table~\ref{tab:joint_table}), yet it is outperformed by LOUDAR in the listening test (see Figure~\ref{fig:mushra_diffvox}), with a statistically significant difference ($p=0.002$). A plausible explanation is that Apollo is not affected by the \ac{ae}, whereas the objective metrics are sensitive to the distortions introduced by this component. This is supported by the fact that reconstructions obtained directly from the \ac{ae}—which can be interpreted as an upper bound for LOUDAR—yield substantially higher metric values than the evaluated methods, suggesting that part of the gap is due to metric bias rather than perceptual quality.

Looking more closely at representation-specific metrics, LOUDAR obtains the lowest scores on Singer-ID embeddings. This indicates that the method preserves speaker identity characteristics more faithfully. More broadly, when compared to the \textit{Latent Regressor} and \textit{\ac{ldm} conditional} baselines, LOUDAR achieves overall comparable performance. It consistently outperforms both methods on AFxRep, CLAP, and Singer-ID metrics, considering both pairwise and distributional distances, although it does not match their performance on FxEnc++. Importantly, these improvements translate to perceptual gains, as LOUDAR significantly surpasses the \textit{\ac{ldm} conditional} model in the listening test ($p<0.001$).

A somewhat counterintuitive result is observed for \emph{SRS}. %
While it scores surprisingly low on effect-based embeddings such as AFxRep and FxEnc++, it achieves the lowest CLAP distance and one of the best listening test scores (see Figure~\ref{fig:mushra_diffvox}).
Its perceptual performance is statistically comparable to LOUDAR ($p=0.433$). 

\emph{RemFX} and \emph{VoiceFixer} exhibit the weakest performance overall, likely due to a mismatch between their training data and our task conditions.
Finally, BUDDy, which operates in the waveform domain, is likewise outperformed by LOUDAR, consistent with the guitar experiment (Sec.~\ref{sec:guitar_exp}) in supporting the benefit of restoring in latent space.

\begin{table*}[t]
\centering
\caption{Objective metrics for both singing voice and guitar restoration experiments.
}
\label{tab:joint_table}
\input{tables/joint_table_v2}
\end{table*}

\vspace{0pt}
\subsection{Singing Voice Distortion Restoration}\label{sec:SVDR}
\vspace{0pt}

\begin{figure}[t]
    \centering
    \includegraphics[width=\columnwidth]{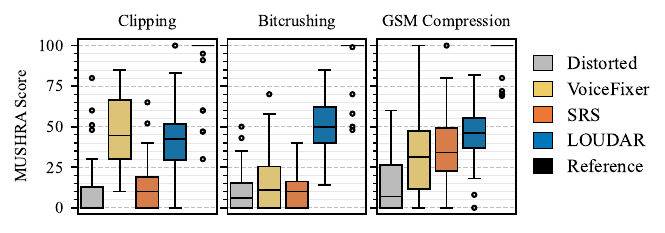}
    \vspace{-20pt}
    \caption{MUSHRA scores for experiment of Sec.~\ref{sec:SVDR}.
\vspace{-25pt}
    } %
    \label{fig:mushra_nhss}
\end{figure}
Additionally, we evaluate LOUDAR on restoring singing voices subjected to three aggressive distortions processed via Pedalboard~\cite{sobot2023Pedalboard} (see Figure~\ref{fig:mushra_nhss}).
To assess generalization across datasets, we evaluate the diffusion prior trained on OpenSinger (Chinese) using English samples from the NHSS dataset \cite{sharma2020nhss}.
The second session of the listening test included 12 such examples (four per distortion).

We compare LOUDAR against two restoration baselines: \emph{SRS} and \emph{VoiceFixer}. As shown in Figure~\ref{fig:mushra_nhss}, LOUDAR exhibits the most stable performance across all conditions. This experiment suggests that \emph{SRS} %
and \emph{VoiceFixer} struggle with distortions potentially underrepresented in their training data. For instance, \emph{SRS} %
performs poorly on clipping and bitcrushing and is outperformed by LOUDAR in GSM compression ($p=0.045$). Similarly, while \emph{VoiceFixer} matches LOUDAR’s scores on clipping ($p=0.191$), 
it is significantly outperformed by LOUDAR in bitcrushing ($p<0.001$) and GSM compression ($p=0.008$).

\vspace{0pt}
\subsection{Guitar Distortion Removal} \label{sec:guitar_exp}
\vspace{0pt}

We further evaluate LOUDAR on dry guitar recovery from amplifier-distorted recordings. 
We train on the GOAT dataset \cite{loth2025goat_paper}, which contains paired DI and effected guitar recordings spanning a wide range of guitar amplifier transformations. For LOUDAR, only the DI recordings are used to train the diffusion prior, whereas the supervised baselines are trained with paired distorted/dry guitar data. For evaluation we use 238 examples from EGDB \cite{chen2022egdb}, balanced across five amplifier conditions.

Following the protocol used for singing voice, we compare LOUDAR against the same supervised baselines, retrained using the paired data in GOAT. 
Additionally,
we evaluate \emph{UnAFx}~\cite{moliner2025unsupervised},
an unsupervised baseline similar to LOUDAR that %
fits a convolutional neural network as a prior, but operates in the waveform domain. All models are trained for a total of 175\,k iterations.

Objective results are reported in Table~\ref{tab:joint_table}.
Across the FxEnc++ and CLAP embeddings, LOUDAR achieves the strongest results.
On AFxRep, LOUDAR remains close to \textit{\ac{ldm} Conditional}, which performs slightly better in KAD and comparably in \ac{cd}.
\emph{Apollo} transfers poorly in this setting.
LOUDAR clearly outperforms \emph{UnAFx},
suggesting that dry-guitar recovery is easier to optimize in latent space than with an unsupervised waveform-domain baseline.

\vspace{0pt}
\section{Conclusions}
\vspace{0pt}

This paper presents LOUDAR, an unsupervised framework for music restoration under arbitrary distortions.
Rather than learning a fixed inverse model for a predefined distortion family,
LOUDAR formulates restoration as joint inference of a~clean latent vector and a latent operator,
regularized by an unconditional \ac{ldm} prior. 
Operating in the latent space of an \ac{ae} makes this inference tractable while preserving broad applicability to complex distortions.
Across our experiments on singing voice and guitar, LOUDAR consistently improves over distorted inputs and compares favorably with strong baselines, 
showing the clearest gains where supervised models struggle to generalize.

Despite the promising results, LOUDAR inherits important limitations from the latent representation on which it relies.
Since restoration is performed entirely in the latent space of an \ac{ae}, the best attainable output quality is bounded by that  reconstruction fidelity of the autoencoder. Moreover, LOUDAR also depends on the encoding of the distorted observation itself: if this is mapped unreliably into latent space, the subsequent restoration can become unpredictable. This limitation becomes more pronounced for signals that are weakly represented by the encoder. 

A second limitation arises from prior mismatch. The diffusion prior may favor samples that are plausible under its training distribution rather than faithfully recovering the original signal, potentially introducing corpus-specific timbral or phonetic artifacts. For instance, heavily distorted English singing can acquire characteristics of the Chinese singing data used to train the prior. Furthermore, LOUDAR is sensitive to strong measurement noise. While it can handle mild interference, performance degrades when noise dominates the observation, as the method is designed to project inputs toward a clean-audio manifold rather than explicitly separate large additive noise components. A natural direction for future work is to combine LOUDAR with discriminative denoising approaches to improve robustness in such conditions.

\newpage

\section{Acknowledgments}
This study was supported by the Czech Science Foundation (Project No. 23-07294S), the Research Council of Finland (Grant no.~371845, REMUS---Restoring Music Recordings Using Generative Models), and the HUCE infrastructure of the Aalto School of Electrical Engineering.
We acknowledge the computational resources from the Aalto University “Science-IT” project. 
We thank all anonymous participants of the listening tests.

\section{Ethics Statement}

All responses from the listening test participants were collected anonymously --- no names or other identifying information were recorded.

\bibliography{ISMIRtemplate}

\end{document}

%% file: figures/latent_operator_figure.tex
\begin{figure}[t]
\centering
\resizebox{0.85\columnwidth}{!}{
\begin{tikzpicture}[
    x=0.78cm,
    y=0.92cm,
    font=\normalsize,
    >=Latex,
    op/.style={draw, rounded corners, fill=gray!15, minimum width=1.05cm, minimum height=0.60cm, align=center},
    enc/.style={draw, rounded corners, dashed, fill=green!8, minimum width=1.00cm, minimum height=0.55cm, align=center},
    cleanwave/.style={blue!75!black, line width=0.8pt, line cap=round},
    distwave/.style={red!70!black, line width=0.8pt, line cap=round},
    addwave/.style={red!55!black, line width=0.45pt, line cap=round, opacity=0.35},
    cleanlatent/.style={draw=blue!75!black, line width=0.35pt},
    distlatent/.style={draw=red!55!black, line width=0.35pt},
    cleantext/.style={text=blue!75!black},
    disttext/.style={text=red!70!black},
    noisetext/.style={text=gray!75!black}
]
    \node[anchor=east, font=\normalsize\bfseries] at (-0.00,0.00) {Waveform};
    \node[anchor=east, font=\normalsize\bfseries] at (-0.35,-2.30) {Latent};

    \node[cleantext] (x) at (0.8,0.0) {$\mathbf{x}$};
    \node[op] (f) at (3.0,0.0) {$f$};
    \node[circle, draw, inner sep=0.6pt] (plusw) at (4.8,0.0) {$+$};
    \node[disttext] (y) at (6.8,0.0) {$\mathbf{y}$};
    \node[noisetext] (n) at (4.8,0.75) {$\mathbf{n}$};

    \node[cleantext] (zx) at (0.8,-1.77) {$\mathbf{z}_0$};
    \node[op] (g) at (3.0,-1.77) {$g_{\boldsymbol{\phi}}$};
    \node[circle, draw, inner sep=0.6pt] (plusz) at (4.8,-1.77) {$+$};
    \node[disttext] (zy) at (6.8,-1.77) {$\mathbf{z}_y$};
    \node[noisetext] (zn) at (4.8,-1.00) {$\mathbf{r}$};

    \node[enc] (ex) at (0.8,-0.88) {$E(\cdot)$};
    \node[enc] (ey) at (6.8,-0.88) {$E(\cdot)$};

    \begin{scope}[shift={(0.8,0.62)}]
        \draw[cleanwave] plot[smooth, domain=-1.15:1.15, samples=140]
            (\x,{0.17*sin(940*\x)});
    \end{scope}

    \begin{scope}[shift={(6.8,0.62)}]
        \draw[red!35!white, dashed, line width=0.65pt]
            plot[smooth, domain=-1.15:1.15, samples=140]
            (\x,{0.13*sin(940*\x)});
        \draw[distwave]
            plot[smooth, domain=-1.15:1.15, samples=180]
            (\x,{0.13*sin(940*\x) + 0.05*sin(2820*\x + 18) + 0.015*cos(1570*\x)});
    \end{scope}

    \begin{scope}[shift={(0.8,-2.50)}]
        \foreach \col/\a/\b/\c in {
            0/18/30/44,
            1/26/40/56,
            2/36/52/68,
            3/48/66/82,
            4/62/80/92,
            5/44/60/76,
            6/28/42/58
        } {
            \pgfmathsetmacro{\xleft}{-1.05 + 0.30*\col}
            \pgfmathsetmacro{\xright}{\xleft + 0.30}
            \fill[blue!\a] (\xleft,-0.45) rectangle (\xright,-0.15);
            \fill[blue!\b] (\xleft,-0.15) rectangle (\xright,0.15);
            \fill[blue!\c] (\xleft,0.15) rectangle (\xright,0.45);
        }
        \draw[cleanlatent] (-1.05,-0.45) rectangle (1.05,0.45);
        \foreach \xgrid in {-0.75,-0.45,-0.15,0.15,0.45,0.75} {
            \draw[cleanlatent] (\xgrid,-0.45) -- (\xgrid,0.45);
        }
        \foreach \ygrid in {-0.15,0.15} {
            \draw[cleanlatent] (-1.05,\ygrid) -- (1.05,\ygrid);
        }
    \end{scope}

    \begin{scope}[shift={(6.8,-2.50)}]
        \foreach \col/\a/\b/\c in {
            0/12/18/26,
            1/16/24/32,
            2/22/30/39,
            3/29/38/48,
            4/38/47/56,
            5/27/36/45,
            6/18/26/34
        } {
            \pgfmathsetmacro{\xleft}{-1.05 + 0.30*\col}
            \pgfmathsetmacro{\xright}{\xleft + 0.30}
            \fill[red!\a] (\xleft,-0.45) rectangle (\xright,-0.15);
            \fill[red!\b] (\xleft,-0.15) rectangle (\xright,0.15);
            \fill[red!\c] (\xleft,0.15) rectangle (\xright,0.45);
        }
        \draw[distlatent] (-1.05,-0.45) rectangle (1.05,0.45);
        \foreach \xgrid in {-0.75,-0.45,-0.15,0.15,0.45,0.75} {
            \draw[distlatent] (\xgrid,-0.45) -- (\xgrid,0.45);
        }
        \foreach \ygrid in {-0.15,0.15} {
            \draw[distlatent] (-1.05,\ygrid) -- (1.05,\ygrid);
        }
    \end{scope}

    \draw[->] (x.east) -- (f.west);
    \draw[->] (f.east) -- (plusw.west);
    \draw[->] (n.south) -- (plusw.north);
    \draw[->] (plusw.east) -- (y.west);

    \draw[->] (x.south) -- (ex.north);
    \draw[->] (ex.south) -- (zx.north);
    \draw[->] (y.south) -- (ey.north);
    \draw[->] (ey.south) -- (zy.north);

    \draw[->] (zx.east) -- (g.west);
    \draw[->] (g.east) -- (plusz.west);
    \draw[->] (zn.south) -- (plusz.north);
    \draw[->] (plusz.east) -- (zy.west);
\end{tikzpicture}
}
\caption{Parallel view of degradation in waveform and latent space. Clean signals are shown in blue and distorted signals in red.
}
\label{fig:latent-operator}
\end{figure}

%% file: figures/inference_figure.tex
\begin{figure}[t]
\centering
\resizebox{0.83\columnwidth}{!}{%
\begin{tikzpicture}[
    x=1.00cm,
    y=0.72cm,
    font=\small,
    >=Latex,
    line join=round,
    flow/.style={->, line width=0.9pt},
    back/.style={->, dashed, line width=0.85pt},
    faint/.style={draw=black!18, line width=0.45pt, rounded corners},
    title/.style={font=\bfseries, fill=white, inner sep=1pt},
    label/.style={align=center, inner sep=1pt},
    statelabel/.style={align=left, inner sep=1pt},
    card/.style={minimum width=2.35cm, minimum height=1.65cm, inner sep=2pt},
    waveobs/.style={card, draw=red!60!black, fill=red!4, rounded corners, line width=0.55pt},
    waveclean/.style={card, draw=blue!65!black, fill=blue!4, rounded corners, line width=0.55pt},
    latnoise/.style={card},
    latstate/.style={card},
    latobs/.style={card},
    latclean/.style={card},
    latnext/.style={card},
    phase/.style={draw, rounded corners, line width=0.55pt, minimum height=0.86cm, inner sep=2pt, align=center},
    encode/.style={phase,  fill=green!10, minimum width=1.45cm},
    estep/.style={phase, fill=violet!12, minimum width=2.75cm},
    mstep/.style={phase, fill=orange!10, minimum width=3.05cm},
    update/.style={phase, fill=green!10, minimum width=3.35cm},
    decode/.style={phase, fill=orange!10, minimum width=1.55cm}
]

\def\latcell{0.13}
\def\lathalf{0.39}
\def\ytop{3.40}
\def\yinit{2.95}
\def\ye{1.35}
\def\ym{-0.40}
\def\yu{-2.50}
\def\yfinal{-4.40}

\coordinate (zinitcell) at (-0.7,\yinit);
\coordinate (ztcell) at (-0.7,\ye);
\coordinate (znextcell) at (-0.7,\yu);
\coordinate (zfinalcell) at (-0.7,\yfinal);
\coordinate (zxglyph) at (4.50,{\ye+0.8});
\coordinate (zyglyph) at (5.80,{\ytop+0.7});
\coordinate (zywest) at ({5.80-\lathalf},\ytop);
\coordinate (zyfeed) at (5.80,{\ytop-\lathalf});
\coordinate (zysplit) at (5.80,0.45);

\node[statelabel, anchor=west] (zinit) at (-0.1,\yinit) {$\mathbf{z}_{\tau_M}$};
\node[statelabel, anchor=west] (zt) at (-0.1,\ye) {$\mathbf{z}_{\tau_i}$};
\node[statelabel, anchor=west] (znext) at (-0.1,\yu) {$\mathbf{z}_{\tau_{i-1}}$};
\node[statelabel, anchor=west] (zfinal) at (-0.1,\yfinal) {$\mathbf{z}_{\tau_0}$};
\node[decode, minimum width=1.30cm] (dec) at (2.90,\yfinal) {Decoder\\$D(\cdot)$};
\node[label] (xhat) at (4.85,\yfinal) {$\hat{\mathbf{x}}$};

\node at (0.2,2.2) {$\vdots$};
\node at (0.2,-3.25) {$\vdots$};

\node[label] (ywave) at (1.25,\ytop) {$\mathbf{y}$};
\node[encode, minimum width=1.30cm, minimum height=0.80cm] (enc) at (3.30,\ytop) {Encoder\\$E(\cdot)$};
\node[label, anchor=south] (zy) at (5.80,{\ytop-0.25}) {$\mathbf{z}_y$};

\node[estep, minimum width=2.10cm, minimum height=0.96cm, text width=1.85cm] (estepbox) at (2.65,\ye) {\textbf{E-step}\\ $\hat{\mathbf{z}}_\theta(\mathbf{z}_{\tau_i}, \tau_i)$};
\node[label, anchor=north] (zxhat) at (4.50,{\ye+0.3}) {$\hat{\mathbf{z}}_0^{(i)}$};
\node[mstep, minimum width=3.25cm, minimum height=0.98cm, text width=3.30cm] (mstepbox) at (3.5,\ym) {\textbf{M-step}\\ fit $g_{\boldsymbol{\phi}}(\hat{\mathbf{z}}_x^{(i)}) + \mathbf{r}$ to $\mathbf{z}_y$};
\node[update, minimum width=2.35cm, minimum height=1.08cm, text width=3.60cm] (updatebox) at (3.5,\yu) {\textbf{Update}\\ $ \mathbf{z}_{\tau_i} - \Delta\tau_i\,[\mathbf{s}_\theta(\mathbf{z}_{\tau_i}, \tau_i) + l(\mathbf{z}_y, \mathbf{z}_{\tau_i}; {\boldsymbol{\phi}}, \mathbf{r})]$};

\begin{scope}[shift={([yshift=0.22cm]ywave.north)}]
    \draw[red!70!black, line width=0.9pt]
        plot[smooth, domain=-0.88:0.88, samples=180]
        (\x,{0.14*sin(940*\x) + 0.05*sin(2850*\x + 17) + 0.02*cos(1510*\x)});
\end{scope}

\begin{scope}[shift={([yshift=0.22cm]xhat.north)}]
    \draw[blue!75!black, line width=0.9pt]
        plot[smooth, domain=-0.88:0.88, samples=160]
        (\x,{0.17*sin(940*\x)});
\end{scope}

\begin{scope}[shift={(zinitcell)}]
    \foreach \row in {0,...,5} {
        \foreach \col in {0,...,5} {
            \pgfmathsetmacro{\xleft}{-\lathalf + \latcell*\col}
            \pgfmathsetmacro{\xright}{\xleft + \latcell}
            \pgfmathsetmacro{\ybottom}{-\lathalf + \latcell*\row}
            \pgfmathsetmacro{\ytop}{\ybottom + \latcell}
            \pgfmathtruncatemacro{\a}{mod(37*\col*\col + 19*\row*\row + 23*\col*\row + 17*\col + 11*\row + 29, 88) + 6}
            \fill[gray!\a] (\xleft,\ybottom) rectangle (\xright,\ytop);
        }
    }
    \draw[gray!70!black, line width=0.36pt] (-\lathalf,-\lathalf) rectangle (\lathalf,\lathalf);
    \foreach \xgrid in {-0.26,-0.13,0.00,0.13,0.26} {\draw[gray!70!black, line width=0.28pt] (\xgrid,-\lathalf) -- (\xgrid,\lathalf);} 
    \foreach \ygrid in {-0.26,-0.13,0.00,0.13,0.26} {\draw[gray!70!black, line width=0.28pt] (-\lathalf,\ygrid) -- (\lathalf,\ygrid);} 
\end{scope}

\begin{scope}[shift={(ztcell)}]
    \foreach \row in {0,...,5} {
        \foreach \col in {0,...,5} {
            \pgfmathsetmacro{\xleft}{-\lathalf + \latcell*\col}
            \pgfmathsetmacro{\xright}{\xleft + \latcell}
            \pgfmathsetmacro{\ybottom}{-\lathalf + \latcell*\row}
            \pgfmathsetmacro{\ytop}{\ybottom + \latcell}
            \pgfmathtruncatemacro{\noisea}{mod(37*\col*\col + 19*\row*\row + 23*\col*\row + 17*\col + 11*\row + 29, 84) + 8}
            \pgfmathsetmacro{\corea}{48*exp(-((\col-4.1)^2 + 1.35*(\row-4.1)^2)/4.9)}
            \pgfmathsetmacro{\coreb}{23*exp(-((\col-1.2)^2 + 1.10*(\row-1.2)^2)/3.5)}
            \pgfmathsetmacro{\ridge}{11*exp(-((\col-4.9)^2 + 0.90*(\row-1.5)^2)/2.7)}
            \pgfmathtruncatemacro{\cleana}{min(82, max(28, round(20 + 0.84*(\corea + \coreb + \ridge))))}
            \fill[gray!\noisea] (\xleft,\ybottom) rectangle (\xright,\ytop);
            \fill[blue!\cleana, opacity=0.28] (\xleft,\ybottom) rectangle (\xright,\ytop);
        }
    }
    \draw[blue!46!black, line width=0.36pt] (-\lathalf,-\lathalf) rectangle (\lathalf,\lathalf);
    \foreach \xgrid in {-0.26,-0.13,0.00,0.13,0.26} {\draw[blue!46!black, line width=0.28pt] (\xgrid,-\lathalf) -- (\xgrid,\lathalf);} 
    \foreach \ygrid in {-0.26,-0.13,0.00,0.13,0.26} {\draw[blue!46!black, line width=0.28pt] (-\lathalf,\ygrid) -- (\lathalf,\ygrid);} 
\end{scope}

\begin{scope}[shift={(znextcell)}]
    \foreach \row in {0,...,5} {
        \foreach \col in {0,...,5} {
            \pgfmathsetmacro{\xleft}{-\lathalf + \latcell*\col}
            \pgfmathsetmacro{\xright}{\xleft + \latcell}
            \pgfmathsetmacro{\ybottom}{-\lathalf + \latcell*\row}
            \pgfmathsetmacro{\ytop}{\ybottom + \latcell}
            \pgfmathtruncatemacro{\noisea}{max(4, round(0.88*(mod(37*\col*\col + 19*\row*\row + 23*\col*\row + 17*\col + 11*\row + 29, 84) + 8)))}
            \pgfmathsetmacro{\corea}{50*exp(-((\col-4.1)^2 + 1.35*(\row-4.1)^2)/4.6)}
            \pgfmathsetmacro{\coreb}{24*exp(-((\col-1.2)^2 + 1.10*(\row-1.2)^2)/3.2)}
            \pgfmathsetmacro{\ridge}{12*exp(-((\col-4.9)^2 + 0.90*(\row-1.5)^2)/2.4)}
            \pgfmathtruncatemacro{\cleana}{min(88, max(32, round(24 + 0.90*(\corea + \coreb + \ridge))))}
            \fill[gray!\noisea] (\xleft,\ybottom) rectangle (\xright,\ytop);
            \fill[blue!\cleana, opacity=0.36] (\xleft,\ybottom) rectangle (\xright,\ytop);
        }
    }
    \draw[blue!56!black, line width=0.36pt] (-\lathalf,-\lathalf) rectangle (\lathalf,\lathalf);
    \foreach \xgrid in {-0.26,-0.13,0.00,0.13,0.26} {\draw[blue!56!black, line width=0.28pt] (\xgrid,-\lathalf) -- (\xgrid,\lathalf);} 
    \foreach \ygrid in {-0.26,-0.13,0.00,0.13,0.26} {\draw[blue!56!black, line width=0.28pt] (-\lathalf,\ygrid) -- (\lathalf,\ygrid);} 
\end{scope}

\begin{scope}[shift={(zfinalcell)}]
    \foreach \row in {0,...,5} {
        \foreach \col in {0,...,5} {
            \pgfmathsetmacro{\xleft}{-\lathalf + \latcell*\col}
            \pgfmathsetmacro{\xright}{\xleft + \latcell}
            \pgfmathsetmacro{\ybottom}{-\lathalf + \latcell*\row}
            \pgfmathsetmacro{\ytop}{\ybottom + \latcell}
            \pgfmathsetmacro{\corea}{74*exp(-((\col-4.0)^2 + 1.6*(\row-4.1)^2)/1.5)}
            \pgfmathsetmacro{\coreb}{40*exp(-((\col-1.1)^2 + 1.2*(\row-1.1)^2)/1.0)}
            \pgfmathsetmacro{\ridge}{26*exp(-((\col-4.9)^2 + 0.8*(\row-1.7)^2)/0.9)}
            \pgfmathtruncatemacro{\a}{min(98, max(14, round(8 + 1.55*(\corea + \coreb + \ridge))))}
            \fill[blue!\a] (\xleft,\ybottom) rectangle (\xright,\ytop);
        }
    }
    \draw[blue!70!black, line width=0.36pt] (-\lathalf,-\lathalf) rectangle (\lathalf,\lathalf);
    \foreach \xgrid in {-0.26,-0.13,0.00,0.13,0.26} {\draw[blue!70!black, line width=0.28pt] (\xgrid,-\lathalf) -- (\xgrid,\lathalf);} 
    \foreach \ygrid in {-0.26,-0.13,0.00,0.13,0.26} {\draw[blue!70!black, line width=0.28pt] (-\lathalf,\ygrid) -- (\lathalf,\ygrid);} 
\end{scope}

\begin{scope}[shift={(zyglyph)}]
    \foreach \row in {0,...,5} {
        \foreach \col in {0,...,5} {
            \pgfmathsetmacro{\xleft}{-\lathalf + \latcell*\col}
            \pgfmathsetmacro{\xright}{\xleft + \latcell}
            \pgfmathsetmacro{\ybottom}{-\lathalf + \latcell*\row}
            \pgfmathsetmacro{\ytop}{\ybottom + \latcell}
            \pgfmathsetmacro{\corea}{60*exp(-((\col-3.9)^2 + 1.5*(\row-4.0)^2)/2.0)}
            \pgfmathsetmacro{\coreb}{32*exp(-((\col-1.3)^2 + 1.2*(\row-1.4)^2)/1.6)}
            \pgfmathsetmacro{\ridge}{18*exp(-((\col-4.8)^2 + 0.9*(\row-1.8)^2)/1.4)}
            \pgfmathtruncatemacro{\perturb}{mod(11*\col + 7*\row + 5*\col*\row + 9, 10)}
            \pgfmathtruncatemacro{\a}{min(84, max(16, round(12 + \corea + \coreb + \ridge + 0.5*\perturb)))}
            \fill[red!\a] (\xleft,\ybottom) rectangle (\xright,\ytop);
        }
    }
    \draw[red!55!black, line width=0.36pt] (-\lathalf,-\lathalf) rectangle (\lathalf,\lathalf);
    \foreach \xgrid in {-0.26,-0.13,0.00,0.13,0.26} {\draw[red!55!black, line width=0.28pt] (\xgrid,-\lathalf) -- (\xgrid,\lathalf);} 
    \foreach \ygrid in {-0.26,-0.13,0.00,0.13,0.26} {\draw[red!55!black, line width=0.28pt] (-\lathalf,\ygrid) -- (\lathalf,\ygrid);} 
\end{scope}

\begin{scope}[shift={(zxglyph)}]
    \foreach \row in {0,...,5} {
        \foreach \col in {0,...,5} {
            \pgfmathsetmacro{\xleft}{-\lathalf + \latcell*\col}
            \pgfmathsetmacro{\xright}{\xleft + \latcell}
            \pgfmathsetmacro{\ybottom}{-\lathalf + \latcell*\row}
            \pgfmathsetmacro{\ytop}{\ybottom + \latcell}
            \pgfmathsetmacro{\corea}{48*exp(-((\col-4.0)^2 + 1.35*(\row-4.0)^2)/4.7)}
            \pgfmathsetmacro{\coreb}{22*exp(-((\col-1.2)^2 + 1.1*(\row-1.2)^2)/3.4)}
            \pgfmathsetmacro{\ridge}{10*exp(-((\col-4.8)^2 + 0.9*(\row-1.5)^2)/2.6)}
            \pgfmathtruncatemacro{\a}{min(80, max(26, round(18 + 0.82*(\corea + \coreb + \ridge))))}
            \fill[blue!\a] (\xleft,\ybottom) rectangle (\xright,\ytop);
        }
    }
    \draw[blue!65!black, line width=0.36pt] (-\lathalf,-\lathalf) rectangle (\lathalf,\lathalf);
    \foreach \xgrid in {-0.26,-0.13,0.00,0.13,0.26} {\draw[blue!65!black, line width=0.28pt] (\xgrid,-\lathalf) -- (\xgrid,\lathalf);} 
    \foreach \ygrid in {-0.26,-0.13,0.00,0.13,0.26} {\draw[blue!65!black, line width=0.28pt] (-\lathalf,\ygrid) -- (\lathalf,\ygrid);} 
\end{scope}

\draw[flow] (ywave.east) -- (enc.west);
\draw[flow] (enc.east) -- (zywest);

\coordinate (ztsplit) at ([xshift=0.12cm]zt.east);
\draw[line width=0.9pt] (zt.east) -- (ztsplit);
\draw[flow] (ztsplit) -- (estepbox.west);
\draw[flow] (estepbox.east) -- (zxhat);
\draw[flow] (zxhat) -- ([xshift=1.01cm]mstepbox.north);
\draw[flow] (zyfeed) |- (mstepbox.east);
\draw[flow] (zyfeed) |- (updatebox.east);
\draw[flow] (ztsplit) |- (updatebox.north west);
\draw[flow] (mstepbox.south) -- (updatebox.north);
\textcolor{red}{poor text only}
\node[label, right, fill=white, inner sep=1pt] at ([xshift=0.10cm,yshift=-0.20cm]mstepbox.south) {${\boldsymbol{\phi}},\mathbf{r}$};
\draw[flow] (updatebox.west) -- (znext.east);

\draw[flow] ([xshift=1.15cm]zfinalcell) -- (dec.west);
\draw[flow] (dec.east) -- (xhat.west);

\end{tikzpicture}%
}
\vspace{-8pt}\caption{Iterative inference procedure. The observed latent $\mathbf{z}_y$ is refined over $M$ E-M update cycles to produce the final estimate $\mathbf{z}_{\tau_0}$, which is decoded into waveform $\hat{\mathbf{x}}$.
}
\label{fig:inference-algorithm}
\end{figure}

%% file: figures/latent_operator_arch_figure.tex
\begin{figure}[t]
\centering
\resizebox{\linewidth}{!}{
\begin{tikzpicture}[
    x=1cm,
    y=1cm,
    font=\normalsize,
    >={Latex[length=1.35mm,width=1.0mm]},
    line join=round,
    flow/.style={->, line width=0.8pt},
    flownoarrow/.style={-, line width=0.8pt},
    op/.style={draw, rounded corners, fill=gray!12, minimum width=1.18cm, minimum height=0.74cm, align=center},
    act/.style={draw, rounded corners, fill=orange!12, minimum width=0.68cm, minimum height=0.38cm, align=center},
    add/.style={circle, draw, fill=green!10, minimum size=0.30cm, inner sep=0pt},
    skip/.style={draw=black, line width=0.90pt, rounded corners=1.8pt, -{Latex[length=1.35mm,width=1.0mm]}},
    cleanlatent/.style={draw=blue!75!black, line width=0.35pt},
    distlatent/.style={draw=red!55!black, line width=0.35pt},
    cleantext/.style={text=blue!75!black},
    disttext/.style={text=red!70!black},
    note/.style={align=center, inner sep=1pt},
    uproj/.style={draw=teal!70!black, fill=cyan!18, line width=0.8pt},
    vproj/.style={draw=violet!75!black, fill=magenta!16, line width=0.8pt},
    windowa/.style={draw=orange!70!red, rounded corners=0.6pt, line width=0.58pt, opacity=0.55},
    windowb/.style={draw=orange!85!red, rounded corners=0.6pt, line width=0.78pt, opacity=0.78},
    windowc/.style={draw=orange!95!red, rounded corners=0.6pt, line width=1.00pt, opacity=0.95},
    repeatbox/.style={draw=black!65, dashed, rounded corners=2pt, line width=0.7pt}
]

    \node[cleantext] at (-0.05,-1.48) {\footnotesize $\hat{\mathbf{z}}_0^{(i)}$};
    \node[disttext] at (9.28,-1.23) {\footnotesize $g_{\boldsymbol{\phi}}(\hat{\mathbf{z}}_0^{(i)})$};

    \begin{scope}[shift={(0.95,-2.5)}]
        \foreach \col/\a/\b/\c in {
            0/18/30/44,
            1/26/40/56,
            2/36/52/68,
            3/48/66/82,
            4/62/80/92,
            5/44/60/76,
            6/28/42/58
        } {
            \pgfmathsetmacro{\xleft}{-0.84 + 0.24*\col}
            \pgfmathsetmacro{\xright}{\xleft + 0.24}
            \fill[blue!\a] (\xleft,-0.36) rectangle (\xright,-0.12);
            \fill[blue!\b] (\xleft,-0.12) rectangle (\xright,0.12);
            \fill[blue!\c] (\xleft,0.12) rectangle (\xright,0.36);
        }
        \draw[cleanlatent] (-0.84,-0.36) rectangle (0.84,0.36);
        \foreach \xgrid in {-0.60,-0.36,-0.12,0.12,0.36,0.60} {
            \draw[cleanlatent] (\xgrid,-0.36) -- (\xgrid,0.36);
        }
        \foreach \ygrid in {-0.12,0.12} {
            \draw[cleanlatent] (-0.84,\ygrid) -- (0.84,\ygrid);
        }
        \draw[windowa] (-0.60,-0.39) rectangle (0.12,0.39);
        \draw[windowb] (-0.36,-0.39) rectangle (0.36,0.39);
        \draw[windowc] (-0.12,-0.39) rectangle (0.60,0.39);
        \draw[cleanlatent, ->] (-0.72,-0.5) -- (0.72,-0.5);
        \node[note, anchor=north] at (0,-0.54) {\scriptsize time $N$};
        \draw[cleanlatent, ->] (-1.0,-0.32) -- (-1.0,0.32);
        \node[note, rotate=90, anchor=south] at (-1.15,0.00) {\scriptsize $C$};
    \end{scope}

    \coordinate (uonewest) at (0.78,-1.55);
    \coordinate (uoneeast) at (1.14,-1.55);
    \filldraw[uproj] (0.78,-1.92) -- (1.14,-1.77) -- (1.14,-1.33) -- (0.78,-1.18) -- cycle;
    \node[text=teal!70!black] at (0.96,-1.55) {\scriptsize $U$};
    \node[op] (convone) at (1.95,-1.55) {\scriptsize Causal\\\scriptsize Conv1D};
    \coordinate (vonewest) at (2.76,-1.55);
    \coordinate (voneeast) at (3.12,-1.55);
    \filldraw[vproj] (2.76,-1.77) -- (3.12,-1.92) -- (3.12,-1.18) -- (2.76,-1.33) -- cycle;
    \node[text=violet!75!black] at (2.94,-1.55) {\scriptsize $V$};
    \node[act] (actone) at (3.95,-1.55) {\scriptsize ReLU};
    \node[add] (addone) at (5.00,-1.55) {$+$};
    \coordinate (utwowest) at (5.84,-1.55);
    \coordinate (utwoeast) at (6.20,-1.55);
    \filldraw[uproj] (5.84,-1.92) -- (6.20,-1.77) -- (6.20,-1.33) -- (5.84,-1.18) -- cycle;
    \node[text=teal!70!black] at (6.02,-1.55) {\scriptsize $U$};
    \node[op] (convtwo) at (7.02,-1.55) {\scriptsize Causal\\\scriptsize Conv1D};
    \coordinate (vtwowest) at (7.83,-1.55);
    \coordinate (vtwoeast) at (8.19,-1.55);
    \filldraw[vproj] (7.83,-1.77) -- (8.19,-1.92) -- (8.19,-1.18) -- (7.83,-1.33) -- cycle;
    \node[text=violet!75!black] at (8.01,-1.55) {\scriptsize $V$};
    \node[add] (addtwo) at (8.55,-1.55) {$+$};

    \draw[repeatbox] (0.28,-1.98) rectangle (5.23,-0.88);
    \node[note, anchor=north west] at (3.42,-2.04) {\scriptsize repeated $\times 2$};

    \draw[flow] (0.3,-1.55) -- (uonewest);
    \draw[flownoarrow] (0.45,-1.55) |- (0.45, -1);
    \draw[flow] (0.45,-1) -| (addone.north);
    \draw[flow] (uoneeast) -- (convone.west);
    \draw[flow] (convone.east) -- (vonewest);
    \draw[flow] (voneeast) -- (actone.west);
    \draw[flow] (actone.east) -- (addone.west);
    \draw[flow] (addone.east) -- (utwowest);
    \draw[flow] (utwoeast) -- (convtwo.west);
    \draw[flow] (convtwo.east) -- (vtwowest);
    \draw[flow] (vtwoeast) -- (addtwo.west);
    \draw[flownoarrow] (5.62,-1.55) |- (5.62, -1);
    \draw[flow] (5.62,-1) -| (addtwo.north);

    \draw[flow] (addtwo.east) -- (8.95,-1.55);

    \begin{scope}[shift={(9.10,-2.5)}]
        \foreach \col/\a/\b/\c in {
            0/14/22/30,
            1/18/28/38,
            2/26/38/50,
            3/35/48/61,
            4/48/61/74,
            5/34/46/58,
            6/22/32/42
        } {
            \pgfmathsetmacro{\xleft}{-0.84 + 0.24*\col}
            \pgfmathsetmacro{\xright}{\xleft + 0.24}
            \fill[red!\a] (\xleft,-0.36) rectangle (\xright,-0.12);
            \fill[red!\b] (\xleft,-0.12) rectangle (\xright,0.12);
            \fill[red!\c] (\xleft,0.12) rectangle (\xright,0.36);
        }
        \draw[distlatent] (-0.84,-0.36) rectangle (0.84,0.36);
        \foreach \xgrid in {-0.60,-0.36,-0.12,0.12,0.36,0.60} {
            \draw[distlatent] (\xgrid,-0.36) -- (\xgrid,0.36);
        }
        \foreach \ygrid in {-0.12,0.12} {
            \draw[distlatent] (-0.84,\ygrid) -- (0.84,\ygrid);
        }
        \draw[distlatent, ->] (-0.72,-0.5) -- (0.72,-0.5);
        \node[note, anchor=north] at (0,-0.54) {\scriptsize time $N$};
        \draw[distlatent, ->] (-1.0,-0.32) -- (-1.0,0.32);
        \node[note, rotate=90, anchor=south] at (-1.15,0.00) {\scriptsize $C$};
    \end{scope}
\end{tikzpicture}
} %
\vspace{-20pt}
\caption{\hspace{0pt}Diagram of the proposed latent operator $g_{{\boldsymbol{\phi}}}$.
\vspace{-10pt}
}
\label{fig:latent-operator-arch}
\end{figure}

%% file: tables/joint_table_v2.tex
\resizebox{\textwidth}{!}{%
\begin{tabular}{@{}ll|cccc|cccc@{}}
\toprule
               &               & \multicolumn{4}{c|}{Pairwise cosine distance $\downarrow$}                        & \multicolumn{4}{c}{Distributional distance (KAD) $\downarrow$} \\
Experiment               & Method        & AFxRep            & FxEnc++           & CLAP              & Singer-ID            & AFxRep     & FxEnc++     & CLAP      & Singer-ID               \\ \midrule
Singing  & Distorted     & $0.369 \pm 0.165$ & $0.097 \pm 0.066$ & $0.147 \pm 0.068$ & $0.465 \pm 0.229$    & 28.83      & 10.29       & 18.59     & 46.24                   \\
voice          & SRS           & $0.497 \pm 0.122$ & $0.080 \pm 0.032$ & $0.067 \pm 0.037$ & $0.059 \pm 0.025$    & 72.02      & 16.22       & 6.49      & 9.76                    \\
               & VoiceFixer    & $0.373 \pm 0.105$ & $0.107 \pm 0.048$ & $0.121 \pm 0.039$ & $0.185 \pm 0.092$    & 37.77      & 16.58       & 14.60     & 21.91                   \\
               & BUDDy         & $0.334 \pm 0.149$ & $0.083 \pm 0.045$ & $0.091 \pm 0.048$ & $0.113 \pm 0.077$    & 23.90      & 11.68       & 7.59      & 7.79                    \\
               & RemFx         & $0.322 \pm 0.154$ & $0.083 \pm 0.056$ & $0.110 \pm 0.052$ & $0.268 \pm 0.175$    & 23.33      & 7.19        & 11.44     & 27.28                   \\
               & LDM conditional     & $0.183 \pm 0.086$ & $0.050 \pm 0.022$ & $0.080 \pm 0.034$ & $0.056 \pm 0.030$    & 15.80      & 5.24        & 9.06      & 5.18                    \\
               & Latent regressor   & $0.179 \pm 0.086$ & $0.053 \pm 0.026$ & $0.090 \pm 0.044$ & $0.065 \pm 0.054$    & 18.21      & 7.42        & 11.18     & 7.20                    \\
               & Apollo        & $0.146 \pm 0.086$ & $0.045 \pm 0.038$ & $0.056 \pm 0.031$ & $0.067 \pm 0.033$    & 5.10       & 1.56        & 3.18      & 9.54                    \\
               & \textbf{LOUDAR (ours)} & $0.184 \pm 0.102$ & $0.063 \pm 0.026$ & $0.076 \pm 0.035$ & $0.048 \pm 0.025$    & 14.36      & 6.04        & 6.81      & 3.45   \\
               & AE reconstruction       & $0.100 \pm 0.068$ & $0.026 \pm 0.012$ & $0.047 \pm 0.021$ & $0.013 \pm 0.006$    & 14.54      & 4.96        & 4.39      & 1.75                    \\ \midrule
Guitar         & Distorted     & $0.713 \pm 0.130$ & $0.381 \pm 0.181$ & $0.242 \pm 0.095$ &  & 56.49      & 33.43       & 43.06     &   \\
               & Apollo        & $0.780 \pm 0.116$ & $0.433 \pm 0.189$ & $0.244 \pm 0.092$ &  & 62.21      & 41.13       & 34.30     &     \\
               & UnAFx         & $0.619 \pm 0.161$ & $0.374 \pm 0.185$ & $0.245 \pm 0.067$ &  & 50.87      & 33.53       & 48.39     &     \\
               & Latent regressor   & $0.549 \pm 0.114$ & $0.323 \pm 0.142$ & $0.242 \pm 0.066$ &  & 39.62      & 22.21       & 35.99     &     \\
               & LDM conditional     & $0.547 \pm 0.125$ & $0.330 \pm 0.142$ & $0.237 \pm 0.069$ &  & 35.26      & 20.69       & 33.40     &     \\
               & \textbf{LOUDAR (ours)} & $0.547 \pm 0.120$ & $0.308 \pm 0.142$ & $0.219 \pm 0.067$ &  & 36.00      & 19.96       & 33.28     &     \\
               & AE reconstruction       & $0.203 \pm 0.059$ & $0.105 \pm 0.063$ & $0.152 \pm 0.056$ &  & 12.70      & 6.95        & 21.78     &     \\ \bottomrule
\end{tabular}
}